
%


\font\bigbold=cmbx12
\font\eightrm=cmr8
\font\sixrm=cmr6
\font\fiverm=cmr5
\font\eightbf=cmbx8
\font\sixbf=cmbx6
\font\fivebf=cmbx5
\font\eighti=cmmi8  \skewchar\eighti='177
\font\sixi=cmmi6    \skewchar\sixi='177
\font\fivei=cmmi5
\font\eightsy=cmsy8 \skewchar\eightsy='60
\font\sixsy=cmsy6   \skewchar\sixsy='60
\font\fivesy=cmsy5
\font\eightit=cmti8
\font\eightsl=cmsl8
\font\eighttt=cmtt8
\font\tenfrak=eufm10
\font\eightfrak=eufm8
\font\sevenfrak=eufm7
\font\fivefrak=eufm5
\font\tenbb=msbm10
\font\eightbb=msbm8
\font\sevenbb=msbm7
\font\fivebb=msbm5
\font\tensmc=cmcsc10


\newfam\bbfam
\textfont\bbfam=\tenbb
\scriptfont\bbfam=\sevenbb
\scriptscriptfont\bbfam=\fivebb
\def\Bbb{\fam\bbfam}

\newfam\frakfam
\textfont\frakfam=\tenfrak
\scriptfont\frakfam=\sevenfrak
\scriptscriptfont\frakfam=\fivefrak
\def\frak{\fam\frakfam}

\def\smc{\tensmc}


\def\eightpoint{%
\textfont0=\eightrm   \scriptfont0=\sixrm
\scriptscriptfont0=\fiverm  \def\rm{\fam0\eightrm}%
\textfont1=\eighti   \scriptfont1=\sixi
\scriptscriptfont1=\fivei  \def\oldstyle{\fam1\eighti}%
\textfont2=\eightsy   \scriptfont2=\sixsy
\scriptscriptfont2=\fivesy
\textfont\itfam=\eightit  \def\it{\fam\itfam\eightit}%
\textfont\slfam=\eightsl  \def\sl{\fam\slfam\eightsl}%
\textfont\ttfam=\eighttt  \def\tt{\fam\ttfam\eighttt}%
\textfont\frakfam=\eightfrak \def\frak{\fam\frakfam\eightfrak}%
\textfont\bbfam=\eightbb  \def\Bbb{\fam\bbfam\eightbb}%
\textfont\bffam=\eightbf   \scriptfont\bffam=\sixbf
\scriptscriptfont\bffam=\fivebf  \def\bf{\fam\bffam\eightbf}%
\abovedisplayskip=9pt plus 2pt minus 6pt
\belowdisplayskip=\abovedisplayskip
\abovedisplayshortskip=0pt plus 2pt
\belowdisplayshortskip=5pt plus2pt minus 3pt
\smallskipamount=2pt plus 1pt minus 1pt
\medskipamount=4pt plus 2pt minus 2pt
\bigskipamount=9pt plus4pt minus 4pt
\setbox\strutbox=\hbox{\vrule height 7pt depth 2pt width 0pt}%
\normalbaselineskip=9pt \normalbaselines
\rm}


\def\pagewidth#1{\hsize= #1}
\def\pageheight#1{\vsize= #1}
\def\hcorrection#1{\advance\hoffset by #1}
\def\vcorrection#1{\advance\voffset by #1}

\newif\iftitlepage   \titlepagetrue               
\newtoks\titlepagefoot     \titlepagefoot={\hfil} 
\newtoks\otherpagesfoot    \otherpagesfoot={\hfil\tenrm\folio\hfil}
\footline={\iftitlepage\the\titlepagefoot\global\titlepagefalse
           \else\the\otherpagesfoot\fi}

\def\abstract#1{{\parindent=30pt\narrower\noindent\eightpoint\openup
2pt #1\par}}


\newcount\notenumber\notenumber=1
\def\note#1
{\unskip\footnote{$^{\the\notenumber}$}
{\eightpoint\openup 1pt #1}
\global\advance\notenumber by 1}


\def\frac#1#2{{#1\over#2}}

\def\tfrac#1#2{{\textstyle{#1\over#2}}}
\def\({\left(}
\def\){\right)}
\def\<{\langle}
\def\>{\rangle}

\def\pmb#1{\setbox0=\hbox{$#1$}%
   \kern-.025em\copy0\kern-\wd0
   \kern.05em\copy0\kern-\wd0
   \kern-.025em\raise.0433em\box0 }


\global\newcount\secno \global\secno=0
\global\newcount\meqno \global\meqno=1
\global\newcount\appno \global\appno=0
\newwrite\eqmac
\def\romappno{\ifcase\appno\or A\or B\or C\or D\or E\or F\or G\or H
\or I\or J\or K\or L\or M\or N\or O\or P\or Q\or R\or S\or T\or U\or
V\or W\or X\or Y\or Z\fi}
\def\eqn#1{
        \ifnum\secno>0
            \eqno(\the\secno.\the\meqno)\xdef#1{\the\secno.\the\meqno}
          \else\ifnum\appno>0
            \eqno({\rm\romappno}.\the\meqno)\xdef#1{{\rm\romappno}.\the\meqno}
          \else
            \eqno(\the\meqno)\xdef#1{\the\meqno}
          \fi
        \fi
\global\advance\meqno by1 }

\global\newcount\refno
\global\refno=1 \newwrite\reffile
\newwrite\refmac
\newlinechar=`\^^J
\def\ref#1#2{\the\refno\nref#1{#2}}
\def\nref#1#2{\xdef#1{\the\refno}
\ifnum\refno=1\immediate\openout\reffile=refs.tmp\fi
\immediate\write\reffile{
     \noexpand\item{[\noexpand#1]\ }#2\noexpand\nobreak.}
     \immediate\write\refmac{\def\noexpand#1{\the\refno}}
   \global\advance\refno by1}
\def\semi{;\hfil\noexpand\break ^^J}
\def\nl{\hfil\noexpand\break ^^J}
\def\refn#1#2{\nref#1{#2}}

\def\jmp#1#2#3{{\it J. Math. Phys.} {\bf {#1}} (19{#2}) #3}

\def\mplA#1#2#3{{\it Mod. Phys. Lett.} {\bf A{#1}} (19{#2}) #3}

\def\np#1#2#3{{\it Nucl. Phys.} {\bf B{#1}} (19{#2}) #3}

\def\ann#1#2#3{{\it Ann. Phys.} {\bf {#1}} (19{#2}) #3}


\pageheight{23cm}
\pagewidth{15.5cm}
\hcorrection{-2.5mm}
\magnification \magstep1
\baselineskip=15.5pt plus 1pt minus 1pt
\parskip=5pt plus 1pt minus 1pt


\def\a{\alpha}
\def\b{\beta}
\def\d{\delta}
\def\e{\varepsilon}
\def\s{\sigma}
\def\w{\wedge}

\def\g{{\frak g}}
\def\r{{\frak r}}
\def\h{{\frak h}}
\def\tf{{\frak t}}
\def\cf{{\frak c}}
\def\sf{{\frak s}}

\def\Z{{\Bbb Z}}
\def\Real{{\Bbb R}}

\def\C{{\Bbb C}}

\def\D{{\cal D}}

\def\ket#1{|#1\rangle}

\def\tr{\mathop{\rm tr}\nolimits}
\def\Tr{\mathop{\rm Tr}\nolimits}

\def\R{R}     

\def\SK{{\rm S}_K}
\def\hi{\frac{i}{\hbar}}


{
\eightpoint

\refn\Mackey
{G.W. Mackey, \lq\lq Induced Representations of Groups
and Quantum Mechanics\rq\rq (Benjamin, New York, 1969)}

\refn\Isham
{C.J. Isham, in \lq\lq Relativity, Groups and Topology
II\rq\rq\   eds. B.S. DeWitt and R. Stora (North-Holland,
Amsterdam, 1984)}

\refn\Landsman
{N.P. Landsman, {\it Rev. Math. Phys.} {\bf 2} (1990)
45, 73; {\bf 4} (1992) 503}

\refn\paper
{D. McMullan and I. Tsutsui, \lq\lq On the emergence
of gauge structures and generalized spin
when quantizing on a coset space",
DIAS-STP-93-14}

\refn\Dirac
{P.A.M. Dirac, \lq\lq Lectures on Quantum
Mechanics\rq\rq\ (Yeshiva, New York, 1964)}

\refn\AM
{R. Abraham and J.E. Marsden, \lq\lq Foundations of Mechanics\rq\rq\
Second Edition (Benjamin/Cummings, Reading, 1978)}

\refn\LandsmanB
{N.P. Landsman,
\lq\lq Rieffel induction as generalized quantum Marsden-Weinstein
reduction", DAMTP-93-22}

\refn\Rajaraman
{See, for example,
R. Rajaraman, \lq\lq Solitons and Instantons\rq\rq\
(North-Holland, Amsterdam, 1982)}

\refn\LL
{N.P. Landsman and N. Linden, \np{365}{91}{121}}

\refn\Koch
{M. Koch, \lq\lq Quantum mechanics on homogeneous configuration
spaces and classical particles
in Yang-Mills fields\rq\rq,
Universit\"at Freiburg preprint THEP 91/19}

\refn\OK
{Y. Ohnuki and S. Kitakado, \mplA{7}{92}{2477}; \jmp{34}{93}{2827}}

\refn\Humphreys
{S.E. Humphreys, \lq\lq Introduction to Lie Algebras
and Representation Theory\rq\rq\
(Springer-Verlag, Berlin, 1972)}

\refn\Fad
{L.D. Faddeev, {\it Theor. Math. Phys.} {\bf 1} (1970) 1}

\refn\Senj
{P. Senjanovic, {\it Ann. Phys.} {\bf 100} (1976) 227}

\refn\KN
{S. Kobayashi and K. Nomizu, \lq\lq Foundations of Differential
Geometry\rq\rq\
Vols.I,II (Interscience, New York, 1969)}

\refn\Fadetal
{A. Alekseev, L. Faddeev and S. Shatashvili, {\it J.
Geom. Phys.} {\bf 5} (1989) 391}

\refn\AS
{A. Alekseev and S. Shatashvili, \np{323}{89}{719}}

\refn\Perelomov
{See, for example,
A.M. Perelomov, \lq\lq Integrable Systems of
Classical Mechanics and Lie Algebras\rq\rq\
(Birkh\"auser Verlag, Basel, 1990)}

\refn\BB
{F.A. Bais and P. Batenburg, \np{269}{86}{363}}

\refn\JK
{T. Jaroszewicz and P.S. Kurzepa, \ann{213}{92}{135}}

\refn\Tanimura
{S. Tanimura, \lq\lq Quantum mechanics on manifolds\rq\rq,
Nagoya Univ. preprint DPNU-93-21}

\refn\Laquer
{H.T. Laquer, in \lq\lq Group actions on manifolds\rq\rq,
{\it Contemp. Math. \bf 36} (AMS, Providence, 1983)}

\refn\Wong
{S.K. Wong, {\it Nuovo Cimento} {\bf 65A} (1970) 689}

\refn\Balachandran
{A.P. Balachandran, G. Marmo, B.-S. Skagerstam and A. Stern,
\lq\lq Gauge Symmetries and Fibre Bundles:
Application to Particle Dynamics\rq\rq,
Lecture Notes in Physics {\bf 188}
(Springer-Verlag, Berlin, 1983)}

\refn\EGH
{T. Eguchi, P.B. Gilkey and A.J. Hanson, {\it Phys. Rep.} {\bf 66}
(1980) 213}

\refn\MT
{D. McMullan and I. Tsutsui, \lq\lq Functional gauge structures in
gauge theories\rq\rq, in preparation}

\refn\NOK
{Y. Ohnuki and S. Kitakado,
\lq\lq Quantum mechanics on a closed loop\rq\rq,
Nagoya Univ. preprint CGNU-93-01}

}


{\baselineskip=12pt
\null
\rightline{PLY-MS-93-04}
\rightline{DIAS-STP-93-21}
\rightline{October 1993}
\vfill
}

{\baselineskip=18pt

\centerline{\bigbold
BPST Instanton and Spin
from Inequivalent Quantizations}
}

\vskip 20pt
\centerline
{\smc David McMullan}

\vskip 5pt
{\baselineskip=13pt
\centerline
{School of Mathematics and Statistics}
\centerline
{University of Plymouth}
\centerline
{Drake Circus, Plymouth, Devon PL4 8AA}
\centerline
{U.K.}
\centerline
{(e-mail: d.mcmullan@plymouth.ac.uk)}

\vskip 10pt
\centerline
{\smc and}

\vskip 10pt
\centerline
{\smc Izumi Tsutsui{\eightrm\baselineskip=10pt\footnote{$^*$}
{Address after 15 November 1993:
Institute for Nuclear Study, University of Tokyo,
Midori-cho, Tanashi-shi, Tokyo 188, Japan.}
}}

\vskip 5pt
\centerline
{Dublin Institute for Advanced Studies}
\centerline
{10 Burlington Road, Dublin 4}
\centerline
{Ireland}
\centerline
{(e-mail: tsutsui@stp.dias.ie)}
}

\vskip 40pt
\abstract{%
{\bf Abstract.}\quad
We present a simple alternative to Mackey's account of the (infinite)
inequivalent
quantizations possible  on a coset space G/H.
Our reformulation is based on the
reduction ${\rm G \rightarrow G/H}$ and
employs a generalized form of
Dirac's approach
to the quantization of constrained systems.
When applied to the four-sphere $S^4 \simeq {\rm Spin(5)/Spin(4)}$,
the inequivalent quantizations induce
relativistic spin and
a background BPST instanton; thus they might provide a natural
account of both of these physical entities.
}
\vfill\eject


There is no unique quantization of any classical system. For example, the
simple configuration space $\Real^3$ is purported to have a unique
quantization courtesy of the celebrated Stone--von~Neumann theorem, but
as shown by Mackey [\Mackey],
this is an artifact of how we view $\Real^3$. If we identify $\Real^3$ with
the coset space E(3)/SO(3), where E(3)  is
the Euclidean group, then there are many quantizations possible labelled by
the irreducible unitary representations of SO(3). This possibility
of different quantizations on $\Real^3$ is not just a mathematical
curiosity but has important physical consequences. Indeed, the non-trivial
quantizations (those not described by the Stone--von~Neumann theorem and
hence not equivalent to the standard Schr\"odinger description)
correspond  to quantum systems on
$\Real^3$  with (non-relativistic) spin.

With this example in mind, when we, for example, analyze Yang-Mills
theory we should expect to no longer just talk about its
specific quantization,
but rather the
different quantum sectors possible and speculate about their
physical significance.  However,
there are serious technical problems in extending the above analysis
to a more general configuration space $Q$, such as that of Yang-Mills
theory, and we are forced
to only consider the standard, Schr\"odinger quantization.
This we feel is an
unsatisfactory state of affairs --- one that can only be resolved
through a better understanding of how best to quantize on non-trivial
configuration spaces.

The extension [\Mackey] of the analysis, discussed above
for the configuration space $\Real^3 \simeq {\rm E(3)/SO(3)}$, to
any configuration spaces $Q$ isomorphic to a coset space\note
{The Lie group G under consideration is either
compact, or locally compact and abelian, or a (semidirect)
product of such groups.}
G/H shows that one can construct many
{\it inequivalent} quantum theories labelled by the irreducible
unitary representations of the closed subgroup H of G,
and that there is no {\it a priori} reason to choose
the trivial one (which is derived from the trivial representation of H and
corresponds to the obvious Schr\"odinger type
quantization in terms of wave functions on G/H) over the
other non-trivial ones.
Mackey's approach, and a number of other approaches
developed from it ({\it e.g.}, [\Isham, \Landsman]),
is based on the induced representations of G, and as such
has the unusual  property that it
deals with wave functions which are {\it vector-valued} rather than
scalar-valued, resulting in a path-integral which is
path-ordered,
describing a transition {\it matrix} rather than an
amplitude.  One aim of this Letter is
to announce that there is an alternative, much simpler, method
for  quantizating on G/H which is free from such vector-valued wave functions
and yet gives the same result
(a fuller account of this is given in a separate paper [\paper]).
The essense of the simplification
lies in the recognition that the induced representations used by
Mackey can be recoverd from a generalization of
Dirac's approach to the
quantization of constrained systems [\Dirac].
The proposed generalization states
that the classical first class constraints which implement the classical
reduction ${\rm G \rightarrow G/H}$ must be allowed to become
\lq anomalous', {\it i.e.,} some of the first class constraints
become second class
in the quantum theory.  (At the classical level the generalized
Dirac approach is hence realized effectively by a Marsden-Weinstein
reduction [\AM, \LandsmanB].)
As a concrete application of these ideas
we shall show that when they are applied
to $Q = S^4$ regarded as Spin(5)/Spin(4), the non-trivial quantizations
induce a background
BPST instanton [\Rajaraman] in addition to (relativistic)
spin.  (Regarding $S^4 \simeq {\rm SO(5)/SO(4)}$ leads to a similar
result [\paper].)
A general framework leading to this result was already
discussed before [\LL, \Koch, \OK]; thus our aim here is to present a
simple (and somewhat detailed) account of how this arises and use it to
reinforce our conviction
that the non-triviality of the quantizations on $Q$
can indeed give rise to physically relevant effects.
More speculatively, we feel that this example hints at a new
role for the BPST instanton  as a probe to the finiteness of
space-time.

The systems we wish to quantize are those describing free (geodesic)
motion on the configuration space $Q \simeq {\rm G/H}$,
with respect to the metric
$g_{\a\b}$ induced from the Killing metric on a semisimple group G.
Classically, this dynamics can be recovered from a
reduction of the free motion on the extended configuration space G,
which we now recapitulate.
The kinematical arena for the Hamiltonian description of dynamics on
the Lie group G is the phase space given by the cotangent bundle
$T^*$G [\AM].  This is actually a
trivial bundle over G and can be identified with ${\rm G} \times
\g$ where $\g$ is the Lie algebra of G, allowing
the pair $(g, R)$, with $g \in {\rm G}$ and $R \in \g$,
to represent a point in the phase space.
As with any cotangent bundle, this phase space comes equipped with a
canonical symplectic 2-form $\omega$ from which the Poisson bracket
between functions can be calculated. In terms of the above
trivialization of $T^*$G,
this symplectic 2-form is given by\note
{
We use
the normalized trace
$\Tr (XY) := - {1\over{c}} \tr (\pi(X) \pi(Y))$,
where $\pi(X)$ is the matrix form assigned to $X \in \g$
in an irreducible representation of $\g$, and
$c$ is a constant
needed to make the trace representation-independent.
In terms of a basis $\{T_m\}$ in $\g$ one has
$X = X^m T_m$ for $X \in \g$, and one can raise or lower
the indices using
$\eta_{mn} := \Tr (T_m T_n)$ and its inverse $\eta^{mn}$ with
$\eta^{ml} \eta_{ln} = \d^m_n$.
}
$$
\omega= d\theta\,,
\qquad\hbox{where}\qquad
\theta := - \Tr \R (g^{-1}dg)\,.
\eqn\symplecticG
$$
Using a matrix representation of the elements of G so that $g$
has matrix elements $g_{ij}$, this symplectic form leads to the
fundamental Poisson bracket,
$$
\{g_{ij}\,, g_{kl}\} = 0\,,
\qquad
\{\R_m\,, g_{ij}\} =(gT_m)_{ij}\,,
\qquad
\{\R_m\,, \R_n\} =f^l_{mn}\R_l\,,
\eqn\pbs
$$
where
$f^l_{mn}$ are the structure constants of G, and
$\R_m:= \Tr(T_m\R)$
are the \lq right-currents' which generate
the right action of G on itself
$g \rightarrow g \tilde g$
for $ \tilde g\in\,$G.
For our Hamiltonian on G we take
$$
H = \tfrac12\Tr \R^2 = \tfrac12\eta^{mn} \R_m \R_n\,.
\eqn\hamonG
$$
The equations of motion derived from (\hamonG) are then
$
\frac{d}{dt}\(g^{-1}\dot g\)=0\,
$
which describe geodesic motion on G.

Since
the right action of the subgroup H on G is generated by the
right currents $\R_i=\Tr(T_i\R)$, where $\{T_i\}$
is a basis of the Lie algebra $\h$ of the subgroup H,
the currents $\R_i$, which
form the algebra $\h$ under Poisson bracket
$\{\R_i\,,\R_j\}=f^k_{ij}\R_k$,
can be used to reduce the
phase space $T^*$G to $T^*$(G/H); namely,
the classical reduction is implemented by
imposing the first class constraints
$$
\R_i=0\,.
\eqn\ccon
$$
Dirac's approach for the quantum reduction then converts (\ccon) to
the conditions imposed on the physical states,
$\widehat \R_i \psi_{\rm phy} = 0$.

The basic idea of our generalized Dirac approach, applied to such
coset spaces, is that the classical
constraints (\ccon) are no longer directly transcribed in
their original form to the quantum theory;
rather, one has to take into account the possible \lq
anomalous' behaviour mentioned earlier and replace (\ccon)
with the effective ones:
$$
\R_i = K_i\,,
\eqn\qcon
$$
where $K_i$ are (at this stage
arbitrary) constants.
The ambiguity in the constants $K_i$ signals
the fact that one will have accordingly (infinitely) many
distinct quantizations, and one can show [\paper]
that they are indeed the
possible quantizations described by Mackey.
In fact, we shall see
later that, due to a consistency at the quantum level, $K_i$
must correspond to the set of intergers that
label the highest weight representation $\chi$ of H.
Note that for $K_i \ne 0$
the constraints
$\phi_i := \R_i - K_i = 0$ are
not first class; they are a mixture of first and second class
constraints since
$$
\{\phi_i\,, \phi_j\} =
f^k_{ij}\,\phi_k + \Tr([T_i,T_j]K) \approx \Tr([T_i,T_j]K)\,,
\eqn\no
$$
where we have introduced $K := K^i T_i$.
In order to isolate the first class subset of (\qcon) we
consider the subalgebra
${\sf}_K := {\rm Ker(ad}_K) \cap \h$
consisting of those elements $X \in \h$ for which $[K, X] = 0$.
For a generic $K$, that is, if $K$ is a {\it regular
semisimple} element in $\h$,
the subalgebra ${\sf}_K$ is precisely the Cartan
subalgebra ${\tf}$ of $\h$
containing $K$ [\Humphreys].
If not, ${\sf}_K$ is
larger than ${\tf}$ and, due to the non-degeneracy of ${\tf}$
with respect to the Killing form, admits the
decomposition ${\sf}_K = {\tf} \oplus {\cf}$ where
${\cf}$ is the orthogonal complement.
Choosing a basis $\{T_s\}$ in ${\sf}_K$,
we see that for
any $T_j \in \h$ we have $\Tr ([T_s, T_j] K) = 0$ and hence
the first class components in (\qcon) are given by
$
\phi_s := \Tr T_s (\R - K)\,.
$
Conversely, from the semisimplicity of $\h$ it follows
that these $\phi_s$ form
the maximal set of the first class components in (\qcon).

We now implement the constraints
in the path-integral framework
using the familiar prescription [\Fad, \Senj].
The easiest way to do this is to add gauge fixing conditions $\xi_s=0$ for
the residual first class subset of constraints $\phi_s = 0$
so that the total set of
constraints $\varphi_k := (\phi_i, \xi_s)$ becomes second class.
With this second class set of constraints the
phase space path-integral reads
$$
Z =
\int \D g \, \D \R \,\d(\varphi_k)\,
{\rm det}^{\frac12}\vert \{ \varphi_k, \varphi_{k'} \} \vert\,
\exp \biggl(\hi\int \theta - \hi \int_0^T dt\, H \biggr)\,,
\eqn\ppi
$$
where $\theta$ is the canonical 1-form in (\symplecticG), that is,
$\int \theta = - \int_0^T dt \, \R_m (g^{-1} \dot g)^m$,
and $H$ is the Hamiltonian (\hamonG).
The path-integral measure in (\ppi) is formally defined from
the volume (Liouville) form of the phase space $\omega^N$
($N = {\rm dim \, G}$) by taking its product over time,
$\D g \,\D \R = \prod_t \omega^N(t)$ where
$\omega^N = \prod_{m = 1}^N  (g^{-1} dg)^m\, d\R_m$; thus
$\D g$ is
a product of the Haar measure of the group G over $t$.

The simplicity of the constraints (\qcon),
which are (at most)
linear in the momentum variables $\R_i$, allows us to
implement them trivially
by integrating over all the momentum variables
$\R_m$.  Indeed, since
the determinant factor in (\ppi) is proportional to
${\rm det}\vert \{ \phi_s\,, \xi_{s'} \} \vert$
on the constrained surface, we can choose
the gauge fixing conditions $\xi_s = 0$ so that the determinant
be independent of $\R_m$ and thereby
carry out the integrations on $\R_m$ at once; the result is
$$
Z = \int \D g \, \d(\xi_s)\,
{\rm det} \vert \{ \phi_s, \xi_{s'} \}
\vert\, \exp\biggl(\hi\int_0^T dt\, L_{\rm tot}\biggr)\,,
\eqn\cpi
$$
where
$$
L_{\rm tot} = \frac 1 2 \Tr (g^{-1} \dot g \vert_{\r})^2
-  \Tr K (g^{-1} \dot g\vert_{\h})\,.
\eqn\lag
$$
In the above
we denoted by $\vert_\h$ (or $\vert_\r$) the projection to the
space $\h$ (or $\r$) in $\g$ defined by the orthogonal decompostion
$\g=\h\oplus\r$ where $\r=\h^\perp$, which is automatically {\it
reductive} [\KN],
{\it i.e.}, $[\h,\r]\subset\r$.  From this it follows that
the first term in the Lagrangian (\lag) is invariant
under $g \rightarrow g\,\tilde h$ for $\tilde h \in {\rm H}$
and hence depends only on G/H.
Clearly,
the effects of non-trivial quantizations are contained in
the second term in (\lag) which is proportional to $K$.
Before examining the effects, we wish to
deduce the restrictions on the parameters in $K$ mentioned
earlier.

To this end, we first observe that under
the transformations
$g \rightarrow g \, s$, for $s \in \SK$
where $\SK$ is the exponential group of ${\sf}_K$,
the total Lagrangian varies as
$$
L_{\rm tot} \longrightarrow L_{\rm tot} + \Delta L_{\rm tot}\,,
\qquad \hbox{where} \qquad
\Delta L_{\rm tot} = -  \Tr K (s^{-1} \dot s)\,.
\eqn\ski
$$
Then, parametrizing
$s = e^{\theta^r T_r} e^{\xi^p T_p}$ where
$\{ T_r \}$ and $\{ T_p \}$ are bases in ${\tf}$ and
${\cf}$, we find
$\Delta L_{\rm tot} = - {d\over{dt}}(K_r \theta^r)$.
Thus the Lagrangian is invariant up to a total time derivative,
which is the residual gauge symmetry at the classical level.
However, if we require the symmetry to persist
at the quantum level (which we must to ensure that the path-integral
(\cpi) is independent of the gauge fixing), then we need to
take into account the contribution from the
boundary in the path-integral.  To examine this explicitly,
consider the transition amplitude from an initial
point $g_0$ at $t = 0$ to a final point $g_1$ at $t = T$.
The sum in the path-integral contains
all possible paths $g(t)$ going from $g(0) = g_0$ to
$g(T) = g_1$, but
to each such path there is a class
of paths related to each other by a
gauge transformation,
$g(t) \rightarrow g(t)\, s(t)$
with $s(0) = s(T) = 1$.
The gauge invariance at the quantum level
requires that the paths within a
gauge equivalent class must contribute
to the sum of the path-integral with the same amplitude, {\it i.e.},
they must have the same phase factor.
Using $H_{\alpha_r}$ in the Chevalley basis\note{
In the complex extension $\h_{\rm c}$ of $\h$,
one can choose
the Chevalley basis $\{H_\alpha, E_{\varphi} \}$ where $\alpha$
are simple roots and $\varphi$ are roots [\Humphreys].
To every dominant weight $\chi$
there exists an irreducible representation
--- highest weight
representation --- of $\h$ where the Cartan elements $H_\alpha$
are diagonal; in particular,
on the states $\ket {\chi, \mu}$
specified by the weights $\mu$ connected to the
the dominant weight $\chi$ (identified as the highest weight
in the representation) their eigenvalues are all integer,
$H_\alpha \ket {\chi, \mu} = \mu(H_\alpha) \, \ket{\chi, \mu}$
with
$\mu(H_\alpha) = 2\mu\cdot\alpha/\vert\alpha\vert^2
\in { \Z}$.
This integral property of $H_\alpha$ leads to the
periodicity
$e^{2 n \pi i H_\alpha} = 1$ for $n \in \Z$
in the exponential mapping defined in
the universal covering group
$\widetilde {\rm H}$ of H.
For a non-simply connected group H the periodicity is
different; it is mulitplied by a factor
determined by the discrete normal subgroup
N of $\widetilde {\rm H}$ for which
H $\simeq \widetilde{\rm H}/{\rm N}$.
For instance, for ${\rm Spin}(n) = \widetilde{{\rm SO}}(n)$
we have $n \in \Z$ but for ${\rm SO}(n) \simeq
{\rm Spin}(n)/\Z_2$ we find $2n \in \Z$.
}
for our basis in $\tf$ as
$T_r = \frac1i H_{\alpha_r}$,
we find from the periodic property
that gauge transformations with
$s(t) = e^{\theta^r(t) T_r}$
that respect the boundary condition satisfy
$\theta^r (T) - \theta^r(0) = 2\pi n_r$,
where $n_r$ are integers.  (More precisely, $n_r$ are integers for
the universal covering group $\widetilde {\rm H}$ of H, but
for a non-simply connected group H they are multiple of integers;
see footnote 3.)
{}From this one sees that the requirement
$e^{\hi\int_0^T dt\, \Delta L_{\rm tot}} = 1$, or
$$
\int_0^T dt \, \Delta L_{\rm tot} = - 2\pi\,
n_r K_r = 2\pi\hbar \times
\hbox{integer}\,,
\eqn\quant
$$
for any class of gauge transformations ({\it i.e.}, for any
$n_r$), is equivalent to
$K_r/\hbar \in  \Z$.
But since any weight can be brought to a dominant weight by
using Weyl reflections, which means that we
can always choose the basis $H_{\alpha_r}$ such that $K_r \geq 0$,
we see that these integer parameters
are precisely associated to the integers which label
the highest weight representations of H:
$$
{1\over\hbar} K_r = \chi(H_{\alpha_r})\,, \qquad {\rm for}
\quad r = 1, \ldots, \hbox{rank}\,{\rm H}\,.
\eqn\ident
$$

Let us next examine the dynamical implications of
the Lagrangian (\lag).
For this, it is convenient to decompose $g$ as
$g = \s \, h$ with
$\s \in {\rm G}$, $h \in {\rm H}$,
where $\s = \s(q)$ is a section ${\rm G/H \mapsto G}$ parametrized
by a set of local coordinates $\{q^\alpha\}$ on G/H.
(Of course, $\s$ must necessarily be singular
unless there exists a global section, and this
is, in fact, the reason why the H-connection $A$
discussed later can be topologically non-trivial.)
Then the Lagrangian (\lag) becomes the sum of three terms
$$
L_{\rm tot} = L_{\rm G/H} + L_{O_K} + L_{\rm int}
     = \frac 1 2 g_{\alpha\beta}(q)\, \dot q^\alpha \dot q^\beta
- \Tr K (h^{-1} \dot h)
- \Tr \bigl(h K h^{-1} A_\alpha(q)\bigr)\, \dot q^\alpha \,,
\eqn\llag
$$
where the metric
$g_{\alpha\beta} = \eta_{ab}\, e_{\,\,\alpha}^a\,
e_{\,\,\beta}^b$ on G/H is given from the vielbein
$e = e^a_{\,\,\alpha}\, dq^\alpha\, T_a := \s^{-1}d\s \vert_\r\,$.
The first term $L_{\rm G/H}$ is just the Lagrangian for a free particle
on G/H, whereas $L_{O_K}$ and $L_{\rm int}$, being
proportional to $K$, deserve separate considerations.

First, we note that $L_{O_K}$
is the first order Lagrangian [\Fadetal, \AS]
for the system defined on the coadjoint orbit
$O_K \simeq {\rm H}/\SK$ of the group H passing through $K$
[\Perelomov].
The natural set of local coordinates of the coadjoint orbit
is given by $S_i := - \Tr (T_i h K h^{-1})$, which
describe the \lq generalized spin' in the sense that
they form
the algebra $\h$
under the Dirac bracket
$\{ S_i\,, S_j\}^* = f_{ij}^k \, S_k$
defined with respect to the second class
constraints $\varphi_k = (\phi_i, \xi_s)$.
Upon quantization, this Dirac bracket
is replaced by the quantum commutator,
$$
[ \widehat S_i\,, \widehat S_j ] =
i\hbar \, f_{ij}^k \, \widehat S_k\,.
\eqn\gspin
$$
Observe, on the other hand, that the change of section,
$\s \rightarrow \s \, \tilde h$, $h \rightarrow  \tilde h^{-1} h$
for $\tilde h \in {\rm H}$,
induces the following transformation on the vielbein,
$$
e = e^a T_a \longrightarrow \tilde h^{-1} e\, \tilde h =
e^a M_a^{\,\,b}(\tilde h) \, T_b \,,
\qquad{\rm with} \quad
M_a^{\,\,b}(\tilde h) := \eta^{bc} \Tr (\tilde h^{-1}
T_a \tilde h \,T_c)\,,
\eqn\rot
$$
where $\{ T_a \}$ is a basis in $\r$, which specifies the
vielbein frame in
the tangent space on the coset space G/H.
Since (\rot) leaves
the metric $g_{\alpha\beta}$ invariant, it is
an SO($n$) ($n = {\rm dim\,(G/H)}$) rotation of the vielbein
frame.  In fact,
on account of the reductive decomposition
the complement $\r$ automatically furnishes
a representation of the group H by the adjoint action (\rot),
producing the SO($n$) frame rotation.
Thus, given a representation of
(\gspin),
one can determine the behaviour (respresentation) of the
generalized spin under the \lq space-time'
frame rotation (\rot).

Second, the term $L_{\rm int}$ in
(\llag) describes the interaction of the so-called
canonical H-connection
[\KN, \BB, \JK]
$A := \s^{-1}d\s \vert_\h = A_\alpha dq^\alpha$
minimally coupled to the particle and the generalized spin.
Note that under the change of section the H-connection transforms as
$A \rightarrow {\tilde h}^{-1}A\,\tilde h
+ {\tilde h}^{-1} d\tilde h$, and as a result
the Lagrangian $L_{\rm tot}$
acquires a formal gauge invariance observed
in the Hamiltonian description by Landsman and Linden [\LL]
(see also [\Koch, \OK, \Tanimura]).
The H-connection is concisely characterized by the fact that
its curvature $F := dA + A \w A$
in the vielbein frame has components
$ F_{ab}^i = - f_{ab}^i $ given precisely by the
structure constants
appearing in $[T_a, T_b] = f^i_{ab} T_i$.
Another important feature is that the H-connection
is actually a solution of the Yang-Mills equation on the coset space
G/H
(for a proof, see [\paper, \Laquer]).

{}From the equations of motion derived from $L_{\rm tot}$,
we find that
the generalized spin $S$ obeys
the covariant constancy equations,
$$
D_t S := {{dS}\over{dt}} + [A_\alpha(q), S] \, \dot q^\alpha = 0\,,
\eqn\emh
$$
whereas the trajectory of the particle is determined by
$$
\ddot q^\alpha +
\Gamma^\alpha_{\beta\gamma}(q)\, \dot q^\beta \dot q^\gamma
- g^{\alpha\beta}(q) \,S_i \, F^i_{\beta\gamma}(q)
\, \dot q^\gamma = 0\,,
\eqn\emq
$$
where $\Gamma^\alpha_{\beta\gamma}$ is the Levi-Civita connection.
Eqs.~(\emh) and (\emq)
are essentially the Wong equations [\Wong, \Balachandran]
under the special, background non-abelian potential (the
H-connection) with the couplings (the parameters in $K$)
in $S$ taking only discrete values (they are quantized).

Let us now apply the above construction to the quantization on
the four-sphere $S^4$ with radius $r$, which we regard as the coset
Spin(5)/Spin(4).  The four-sphere $S^4$
may be thought of as a finite Euclidean version of the
four dimensional Minkowski space-time (with $t$ being a proper
time), and our aim is to see the
possible effects of inequivalent quantizations for finite $r$.
We shall use the defining representation of spin(5), {\it i.e.},
the spinor representation of so(5), and choose our bases
in $\h$ and $\r$ as
$$
T_i = \cases{
{1\over{2i}}
\left(\matrix{
\s_i    &  \cr
     &  \s_i  \cr}
\right),
& for $i = 1, \, 2, \, 3$, \cr
{1\over{2i}}
\left(\matrix{
\s_{i-3}    &  \cr
     &  - \s_{i-3}  \cr}
\right),
& for $i = 4, \, 5, \, 6$, \cr
}
\eqn\hrep
$$
and
$$
T_a = \cases{
{1\over{2i}}
\left(\matrix{
       & \s_a \cr
  \s_a &      \cr}
\right),
& for $a = 1, \, 2, \, 3$, \cr
{1\over2}
\left(\matrix{
  & - 1 \cr
1 &   \cr}
\right),
& for $a = 4$, \cr
}
\eqn\no
$$
for which $\tr (T_m\,T_n) = - \d_{mn}$.
To assign the radius $r$ to $S^4$
for our spinor representation we choose the constant
in our \lq $\Tr$' as $c = \frac{1}{r^2}$
(see Appendix C in [\paper]), which leads to
the metric in the vielbein frame $\eta_{ab} = r^2 \d_{ab}$.

Since the two so(4) spinor representations given in (\hrep)
are reducible
(which of course is due to the direct sum structure
so(4) $=$ su(2) $\oplus$ su(2)),
it is already clear that we
have two su(2)-valued variables for our generalized spin.
We shall label the two su(2)
by su$(2)^+$ and su$(2)^-$ and introduce
the chiral basis
$$
T^{+}_i = \frac12 (T_i + T_{i+3}) =
{1\over{2i}}
\left(\matrix{
\s_i    &  \cr
     &  0  \cr}
\right),
\qquad
T^{-}_i = \frac12 (T_i - T_{i+3}) =
{1\over{2i}}
\left(\matrix{
0    &  \cr
     &  \s_i  \cr}
\right),
\eqn\cbasis
$$
for $i = 1, \, 2, \, 3$, and thereby write
$h = h^+\, h^-$
where $h^\pm$ belong to the exponential groups generated by the chiral
su$(2)^\pm$ in the basis (\cbasis).
Setting $K = j^+ T^+_3 + j^- T^-_3$, we find that
the Lagrangian for the coadjoint orbit of Spin(4) consists of those for
the coadjoint orbits of the two ${\rm SU(2)}^\pm$,
{\it i.e.}, for the two conventional spins,
$$
L_{O_K} = L_{\rm spin}^+ + L_{\rm spin}^-\,, \qquad {\rm
where}\qquad
L^\pm_{\rm spin} := - \hbar
j^\pm \Tr T^\pm_3 \bigl((h^\pm)^{-1} \dot h^\pm\bigr)\,.
\eqn\no
$$
Indeed, from (\gspin) we have the spin variables
$S_i^{\pm} = - j^\pm \Tr \bigl( T^\pm_i h^{\pm} T_3^\pm
(h^{\pm})^{-1} \bigr)$ forming
two commuting su(2) algebras upon quantization,
$$
[ \widehat S^+_i\,, \widehat S^+_j ]
= i \hbar \, \e_{ijk}\, \widehat S^+_k\,, \qquad
[ \widehat S^-_i\,, \widehat S^-_j ]
= i \hbar \, \e_{ijk}\, \widehat S^-_k\,, \qquad
[ \widehat S^+_i\,, \widehat S^-_j ] = 0\,.
\eqn\chspin
$$
Now, from the relations,
$$
[T^\pm_i\,, T_a ] = \frac12 \e_{iab} \, T_b  \pm \frac12 \d_{ia}\,
T_4\,, \qquad
[T^\pm_i\,, T_4 ] = \mp \frac12 \d_{ia}\, T_a\,,
\eqn\no
$$
it is also easy to see that the basis
$\{T_{\pm\pm} := T_1 \mp i T_2,\,
T_{\pm\mp} := T_3 \mp i T_4\}$ in the space $\r$ forms
a tensor product representation
$2^+ \otimes 2^-$
of spin $\frac12$ with respect to each chiral su$(2)^\pm$.
Accordingly, the adjoint action (\rot) of H
amounts to the product ${\rm SU(2)^+ \times SU(2)^-}$
transformations in the vielbein frame, which
are \lq double-valued' in terms of SO(4) frame rotations.
This in turn implies that the representations of (\chspin)
determine the spin of the particle in exactly the same manner
as in the Minkowski space case,
where the Lorentz frame rotation (that is, the action of the
proper, orthochronous Lorents group) is realized by
the group action of SL(2,$\C$) consisting of two
chiral SU(2) actions.
Thus we have recovered the conventional,
two su(2) chiral
\lq relativistic' spins from the inequivalent quantizations on
$S^4$.

Turning to the H-connection, we observe that
the su(2)-valued
H-connections $A^\pm$, defined by the decomposition
$A = A^+ + A^-$ in terms of the chiral basis (\cbasis),
couple to the two ${\rm su(2)^\pm}$ spins chirally,
$$
L_{\rm int} = \Tr (S^+ A^+_\alpha) \, \dot q^\alpha
              + \Tr (S^- A^-_\alpha) \, \dot q^\alpha\,.
\eqn\no
$$
It is then easy to confirm that
these chiral H-connections $A^\pm$
are nothing but a BPST instanton and anti-instanton, respectively
[\BB].  We shall here compute explicitly
the Chern number $C_2$ [\EGH] of each of the su(2)-valued
H-connection $A^\pm$.  We start with
$$
F^\pm \w F^\pm = \frac{1} {4r^4}
\e^{abcd}\, f_{ab}^i\, f_{cd}^j \,
T^\pm_i \, T^\pm_j \, \Omega\,,
\eqn\ff
$$
where $F^{\pm}$ are the curvatures corresponding to $A^\pm$ and
$\Omega = r^4 e^1 \w e^2 \w e^3 \w e^4$ is the volume form on $S^4$,
that is, $\int_{S^4} \Omega = {{8 \pi^2 r^4}\over 3}$.
Using the structure constants found in the commutation relations,
$[T_a, T_b] = \e_{abi}\, T_i$ and $[T_a, T_4] = \d_{a,i-3}\, T_i$
for $a$, $b = 1$,  2, 3, one finds that
the instanton number
for $F^\pm$ ---
given by $(-1)$ times $C_2$ ---
is
$$
n^\pm = - C_2 [ F^\pm ] =
- {1\over{8\pi^2}}\int_{S^4} \tr (F^\pm \w F^\pm)
= - {1\over{8\pi^2}}\cdot
  {{8\pi^2 r^4}\over 3}\cdot{{(\mp 3)}\over{r^4}} = \pm 1\,,
\eqn\no
$$
which shows that $A^\pm$ is indeed a BPST instanton
(anti-instanton).  One can also confirm that $F^\pm$
satisfies the self dual (anti-self dual) equation,
$* F^\pm = \pm F^\pm$.

Thus we have seen that,
on the four-sphere $S^4$, a background BPST instanton
(and anti-instanton)
emerges naturally together with relativistic spin.
The instanton effect is in principle observable if
the radius $r$ of the (Euclideanized) space-time is finite,
although the order $F_{\alpha\beta}^i \sim {\cal O}(r^{-2})$
implies that the effect is small for a large $r$, a result
consistent
with the fact that there emerges no such
background potential if we quantize on
$Q = {\Real}^4 \simeq {\rm E(4)/SO(4)}$
instead of $S^4$.
Finally, we wish to mention that the generalized Dirac approach
admits an extension to field theory [\MT], where, for instance,
the inequivalent quantizations in Yang-Mills theory
lead to the $\theta$-term in four dimensions [\Isham]
and to the Chern-Simons term in three dimensions.
Another direction of
extension is to go beyond coset spaces, where similar effects
are expected to occur in general
depending perhaps only on the topology of $Q$
[\Isham, \NOK].

\bigskip
\noindent
{\bf Acknowledgements:}  We both wish to thank John Lewis
for his support in this work, and I.T.\thinspace wishes
to thank
L\'aszl\'o Feh\'er and Lochlainn O'Raifeartaigh for
discussions.

  \vfill\eject\immediate\closeout\reffile
  \centerline{{\bf References}}\bigskip\eightpoint\frenchspacing%
  \input refs.tmp\vfill\eject\nonfrenchspacing

\bye